\documentclass[12 pt,twoside,aps,prd,amsmath,amssymb,
tightenlines,showpacs,showkeys,eqsecnum]{revtex4-1}
\usepackage{mathptmx}
\usepackage{bm}
\usepackage{graphicx}

\usepackage{hyperref}
\newcommand {\G}{\Gamma}

\def \myfigures #1#2#3#4#5#6#7#8
{\begin{figure}[ht]
    \begin{center}
        \includegraphics[width=#2 \textwidth]{#1.eps}
        \hfill
        \includegraphics[width=#6 \textwidth]{#5.eps}
        \parbox[t]{#4\textwidth}{\caption {#3}\label{#1}}
        \hfill
        \parbox[t]{#8\textwidth}{\caption {#7}\label{#5}}
    \end{center}
\end{figure} }

\def\myfigure #1#2#3#4
{\begin{figure}[ht]\begin{center}
\includegraphics[width=#2 \textwidth]{#1.jpg}
\parbox[t]{#4\textwidth}{\caption{#3}\label{#1}}
\end{center}\end{figure}}

\begin{document}
\baselineskip -24pt
\title{Bianchi type-VI anisotropic dark energy model with varying EoS parameter}
\author{Bijan Saha}
\affiliation{Laboratory of Information Technologies\\
Joint Institute for Nuclear Research, Dubna  \\ 141980 Dubna, Moscow
region, Russia} \email{bijan@jinr.ru}
\homepage{http://bijansaha.narod.ru}

\begin{abstract}

Within the scope of an anisotropic Bianchi type-VI cosmological
model we have studied the evolution of the universe filled with
perfect fluid and dark energy. To get the deterministic model of
Universe, we assume that the shear scalar $(\sigma)$ in the model is
proportional to expansion scalar $(\vartheta)$. This assumption
allows only isotropic distribution of fluid. Exact solution to the
corresponding equations are obtained. The EoS parameter for dark
energy as well as deceleration parameter is found to be the time
varying functions. Using the observational data qualitative picture
of the evolution of the universe corresponding to different of its
stages is given. The stability of the solutions obtained is also
studied.
\end{abstract}



\maketitle

\bigskip

\section{Introduction}

After the discovery of late time accelerating mode of expansion of
the Universe a number of models are offered to explain this
phenomenon. Most of the dark energy models such as cosmological
constant, quintessence, Chaplygin gas etc. are modeled with a
constant EoS parameter. Recently in a number of papers different
cosmological models with time dependent EoS parameter was studied
\cite{APSAPSSbi0,APSCPL,PASIJTPbi,PASAPSS,SAPAPSS,YSAPSSbi}. The aim
of the current paper is to extend that study for a Bianchi type-VI
cosmological model.

\section{Basic equations}

A Bianchi type-VI model describes an anisotropic but homogeneous
Universe. This model was studied by several authors
\cite{BVI,GCBVI2010,svBVI,Hoogen,Socorro,Weaver1}, specially due to
the existence of magnetic fields in galaxies which was proved by a
number of astrophysical observations.

Bianchi type-VI model given be given by \cite{BVI,GCBVI2010}
\begin{equation}
ds^2 = dt^2 - a_1^{2} e^{-2mz}\,dx^{2} - a_2^{2} e^{2nz}\,dy^{2} -
a_3^{2}\,dz^2, \label{bvi}
\end{equation}
with $a_1,\,a_2,\,a_3$ being the functions of time only. Here
$m,\,n$ are some arbitrary constants and the velocity of light is
taken to be unity. The metric \eqref{bvi} is known as Bianchi
type-VI model. A suitable choice of $m,\,n$ as well as the metric
functions $a_1,\,a_2,\,a_3$ in the BVI given by \eqref{bvi} evokes
Bianchi-type VI$_0$, V, III, I and FRW  universes.

Here we consider the case when the energy momentum tensor has only
non-trivial diagonal elements, i.e.
\begin{eqnarray}
T_\alpha^\beta = {\rm diag}[T_0^0,\,T_1^1,\,T_2^2,\,T_3^3]
\label{emt}
\end{eqnarray}

Einstein field equations for the metric \eqref{bvi} on account of
\eqref{emt} have the form \cite{BVI}

\begin{subequations}
\label{einbvi}
\begin{eqnarray}
\frac{\ddot a_2}{a_2} +\frac{\ddot a_3}{a_3} +\frac{\dot
a_2}{a_2}\frac{\dot
a_3}{a_3} - \frac{n^2}{a_3^2} &=& \kappa T_{1}^{1}, \label{11bvi}\\
\frac{\ddot a_3}{a_3} +\frac{\ddot a_1}{a_1} +\frac{\dot
a_3}{a_3}\frac{\dot
a_1}{a_1} - \frac{m^2}{a_3^2} &=& \kappa T_{2}^{2}, \label{22bvi} \\
\frac{\ddot a_1}{a_1} +\frac{\ddot a_2}{a_2} +\frac{\dot
a_1}{a_1}\frac{\dot
a_2}{a_2} + \frac{m n}{a_3^2} &=& \kappa T_{3}^{3}, \label{33bvi}\\
\frac{\dot a_1}{a_1}\frac{\dot a_2}{a_2} +\frac{\dot a_2}{a_2}
\frac{\dot a_3}{a_3} + \frac{\dot a_3}{a_3}\frac{\dot a_1}{a_1} -
\frac{m^2 - m n + n^2}{a_3^2} &=&
\kappa T_{0}^{0}, \label{00bvi}\\
m \frac{\dot a_1}{a_1} - n \frac{\dot a_2}{a_2} - (m - n) \frac{\dot
a_3}{a_3} &=& 0. \label{03bvi}
\end{eqnarray}
\end{subequations}

We define the spatial volume of the model \eqref{bvi} as
\begin{equation}
V = a_1 a_2 a_3, \label{Vdef}
\end{equation}
and the average scale factor as
\begin{equation}
a = V^{1/3} = (a_1 a_2 a_3)^{1/3}. \label{adef}
\end{equation}

Let us now find expansion and shear for BVI metric. The expansion is
given by
\begin{equation}
\vartheta = u^\mu_{;\mu} = u^\mu_{\mu} + \G^\mu_{\mu\alpha}
u^\alpha, \label{expansion}
\end{equation}
and the shear is given by
\begin{equation}
\sigma^2 = \frac{1}{2} \sigma_{\mu\nu} \sigma^{\mu\nu},
\label{shear}
\end{equation}
with
\begin{equation}
\sigma_{\mu\nu} = \frac{1}{2}\bigl[u_{\mu;\alpha} P^\alpha_\nu +
u_{\nu;\alpha} P^\alpha_\mu \bigr] - \frac{1}{3} \vartheta
P_{\mu\nu}, \label{shearcomp}
\end{equation}
where the projection vector $P$:
\begin{equation}
P^2 = P, \quad P_{\mu\nu} = g_{\mu\nu} - u_\mu u_\nu, \quad
P^\mu_\nu = \delta^\mu_\nu - u^\mu u_\nu. \label{proj}
\end{equation}
In comoving system we have $u^\mu = (1,0,0,0)$. In this case one
finds
\begin{equation}
\vartheta = \frac{\dot a_1}{a_1} + \frac{\dot a_2}{a_2} + \frac{\dot
a_3}{a_3} = \frac{\dot V}{V}, \label{expbvi}
\end{equation}
and
\begin{eqnarray}
\sigma_{1}^{1} &=& \frac{1}{3}\Bigl(-2\frac{\dot a_1}{a_1} +
\frac{\dot
a_2}{a_2} + \frac{\dot a_3}{a_3}\Bigr) =  \frac{\dot a_1}{a_1} - \frac{1}{3} \vartheta, \label{sh11}\\
\sigma_{2}^{2} &=& \frac{1}{3}\Bigl(-2\frac{\dot a_2}{a_2} +
\frac{\dot a_3}{a_3} +
\frac{\dot a_1}{a_1}\Bigr) =  \frac{\dot a_2}{a_2} - \frac{1}{3} \vartheta, \label{sh22}\\
\sigma_{3}^{3} &=& \frac{1}{3}\Bigl(-2\frac{\dot a_3}{a_3} +
\frac{\dot a_1}{a_1} + \frac{\dot a_2}{a_2}\Bigr) =  \frac{\dot
a_3}{a_3} - \frac{1}{3} \vartheta. \label{sh33}
\end{eqnarray}
One then finds
\begin{equation}
\sigma^ 2 = \frac{1}{2}\biggl[\sum_{i=1}^3 \biggl(\frac{\dot
a_i}{a_i}\biggr)^2 - \frac{1}{3}\vartheta^2\biggr] =
\frac{1}{2}\biggl[\sum_{i=1}^3 H_i^2 -
\frac{1}{3}\vartheta^2\biggr]. \label{sheargen}
\end{equation}

As one sees, neither the expansion nor the components of shear
tensor depend on $m$ or $n$., hence the Bianchi cosmological models
of type VI, VI$_0$, V, III and I has the same expansion and shear
tensor.

The Hubble constant of the model is defined by
\begin{equation}
H = \frac{\dot a}{a} = \frac{1}{3} \Bigl(\frac{\dot a_1}{a_1} +
\frac{\dot a_2}{a_2} + \frac{\dot a_3}{a_3}\Bigr) = \frac{1}{3}
\frac{\dot V}{V}. \label{Hubblebvi}
\end{equation}
The deceleration parameter $q$, and the average anisotropy parameter
$A_m$ are defined by
\begin{eqnarray}
q &=& - \frac{a \ddot a}{\dot a^2} = 2 - 3\frac{V \ddot V}{\dot V^2} , \label{decparvi}\\
A_m &=& \frac{1}{3}\sum_{i=1}^3 \Bigl(\frac{H_i}{H} - 1\Bigr)^2,
\label{anisvi}
\end{eqnarray}
where $H_i$ are the directional Hubble constants:
\begin{equation}
H_1 = \frac{\dot a_1}{a_1}, \quad H_2 = \frac{\dot a_2}{a_2}, \quad
H_3 = \frac{\dot a_3}{a_3}. \label{dirhubc}
\end{equation}

Note that, none of the above defined quantity depends on $m$ or $n$,
hence will be valid for not only BVI, but also for BVI$_0$, BV, BIII
and BI.

\section{Solution to the field equations}

From \eqref{03bvi} immediately follows
\begin{equation}
\bigl(\frac{a_1}{a_3}\bigr)^m = k_1 \bigl(\frac{a_2}{a_3}\bigr)^n,
\quad k_1 = {\rm const.} \label{abcrel}
\end{equation}

We also impose use the proportionality condition, widely used in
literature. Demanding that the expansion is proportion to a
component of the shear tensor, namely
\begin{equation}
\vartheta = N_3 \sigma_3^3.\label{propconvi}
\end{equation}
The motivation behind assuming this condition is explained with
reference to  Thorne \cite{thorne67}, the observations of the
velocity-red-shift relation for extragalactic sources suggest that
Hubble expansion of the universe is isotropic today within $\approx
30$ per cent \cite{kans66,ks66}. To put more precisely, red-shift
studies place the limit
\begin{equation}
\frac{\sigma}{H} \leq 0.3, \label{propconviexp}
\end{equation}
on the ratio of shear $\sigma$ to Hubble constant $H$ in the
neighborhood of our Galaxy today. Collins et al. (1980) have pointed
out that for spatially homogeneous metric, the normal congruence to
the homogeneous expansion satisfies that the condition
$\frac{\sigma}{\theta}$ is constant.

On account of \eqref{expbvi} and \eqref{sh33} we find
\begin{equation}
a_3 = N_0 V^{ \frac{1}{3} + \frac{1}{N_3}}, \quad N_0 = {\rm const.}
\label{a3vi}
\end{equation}

In view of \eqref{Vdef} and \eqref{a3vi} from \eqref{abcrel}  we
find
\begin{eqnarray}
a_1 &=& k_1^{\frac{1}{m+n}} N_0^{\frac{m-2n}{m+n}} V^{\frac{1}{3} + \frac{m - 2n}{3N_3(m+n)}}, \label{a1vi}\\
a_2 &=& k_1^{-\frac{1}{m+n}} N_0^{\frac{n-2m}{m+n}} V^{\frac{1}{3} +
\frac{n - 2m}{3N_3(m+n)}}. \label{a2vi}
\end{eqnarray}

Thus, we have derived metric functions in terms of $V$. In order to
find the equation for $V$ we take the following steps. Subtractions
of \eqref{11bvi} from \eqref{22bvi}, \eqref{33bvi} from
\eqref{33bvi}, and \eqref{33bvi} from \eqref{11bvi} on account of
\eqref{a1vi}, \eqref{a2vi} and \eqref{a3vi} give
\begin{subequations}
\label{V123}
\begin{eqnarray}
\frac{\ddot V}{V} - \frac{N_3 (m+n)^2}{3 N_0^2 V^{2/3+2/N_3}} &=&
\kappa \frac{T_2^2 - T_1^1}{X_{12}}, \label{V12}\\
\frac{\ddot V}{V} - \frac{N_3 (m+n)^2}{3 N_0^2 V^{2/3+2/N_3}} &=&
\kappa \frac{T_3^3 - T_2^2}{X_{23}}, \label{V23}\\
\frac{\ddot V}{V} - \frac{N_3 (m+n)^2}{3 N_0^2 V^{2/3+2/N_3}} &=&
\kappa \frac{T_1^1 - T_3^3}{X_{31}}, \label{V31}\\
\end{eqnarray}
\end{subequations}
where $X_{12} = 3(m-n)/N_3(m+n)$, $X_{23} = - 3m/N_3 (m+n)$ and
$X_{31} = 3n/N_3 (m+n)$. From \eqref{V123} immediately follows
\begin{equation}
\frac{T_2^2 - T_1^1}{X_{12}} = \frac{T_3^3 - T_2^2}{X_{23}} =
\frac{T_1^1 - T_3^3}{X_{31}}. \label{T123}
\end{equation}
After a little manipulation, it could be established that
\begin{equation}
T_1^1 = T_2^2 = T_3^3 \equiv - p. \label{T123iso}
\end{equation}
Thus we conclude that under the proportionality condition, the
energy-momentum distribution of the model should be strictly
isotropic.

Let us now gp back to the equation for $V$ that now reads
\begin{equation}
\ddot V - A_0 V^{(N_3 - 6)/3N_3} = 0, \quad A_0 = \frac{N_3
(m+n)^2}{3 N_0^2}, \label{Vnew}
\end{equation}
which allows the solution in quadrature
\begin{equation}
\int\frac{dV}{\sqrt{A_1 V^{(4N_3-6)/3N_3} + C_0}} = t + t_0, \quad
A_1 =  \frac{3N_3 A_0}{(2N_3 - 3)}, \quad t_0 = {\rm const.}
\label{quadrvi}
\end{equation}

Thus we have the solution to the corresponding equation in
quadrature.

\begin{figure}[ht]
\centering
\includegraphics[height=70mm]{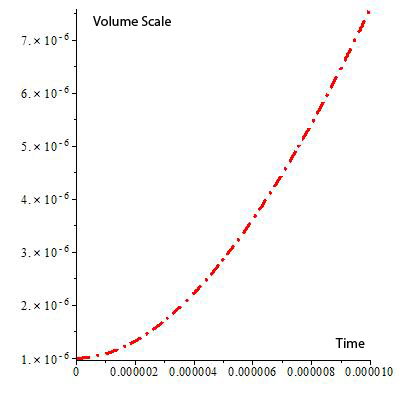} \\
\caption{Evolution of the Universe given by a BVI cosmological
model. } \label{VE-6}.
\end{figure}

Fig. [\ref{VE-6}] shows the evolution of the Universe. As one sees,
it is an expanding one.

\section{Physical aspects of Dark energy model}

Let us now find the expressions for physical quantities.

Inserting \eqref{quadrvi} into \eqref{Hubblebvi} and
\eqref{decparvi} one finds the expression for expansion $\vartheta$,
Hubble parameter $H$:
\begin{equation}
\vartheta = 3H = \sqrt{A_1 V^{-(2N_3+6)/3N_3}+ C_0/V^2},
\label{Hubblevi1}
\end{equation}
and deceleration parameter
\begin{equation}
q = -  \frac{A_0 V^{-(2N_3+6)/3N_3}}{A_1 V^{-(2N_3+6)/3N_3}+
C_0/V^2}. \label{dp1}
\end{equation}
The anisotropy parameter $A_m$ has the expression
\begin{equation}
A_m = \frac{54 (m^2 - mn + n^2)}{N_3^2 (m+n)^2}. \label{aniso}
\end{equation}
The directional Hubble parameters are
\begin{equation}
H_1 = \biggl[\frac{1}{3} - \frac{2n - m}{N_3(m+n)}\biggr]\frac{\dot
V}{V}, \quad H_2 = \biggl[\frac{1}{3} - \frac{2m -
n}{N_3(m+n)}\biggr]\frac{\dot V}{V}, \quad H_3 = \biggl[\frac{1}{3}
+ \frac{1}{N_3}\biggr]\frac{\dot V}{V}. \label{dirhubex}
\end{equation}
\begin{figure}[ht]
\centering
\includegraphics[height=70mm]{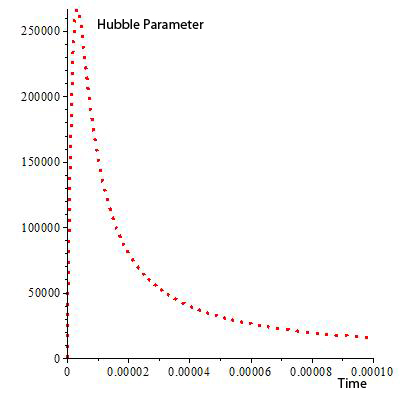} \\
\caption{Evolution of the Hubble parameter } \label{HE-6}.
\end{figure}

\begin{figure}[ht]
\centering
\includegraphics[height=70mm]{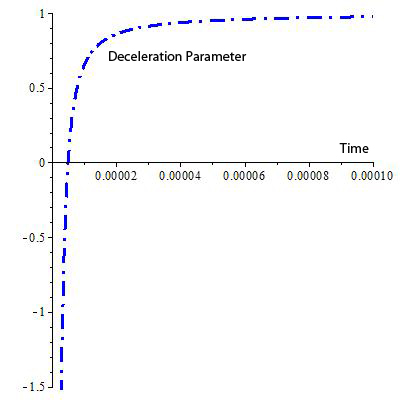} \\
\caption{Evolution of the deceleration parameter. } \label{qE-6}.
\end{figure}

Figs. [\ref{HE-6}] and  [\ref{qE-6}] show the behavior of the Hubble
parameter and deceleration parameter, respectively.

From \eqref{00bvi} we find the expression for energy density
\begin{equation}
\varepsilon = T_0^0 = \frac{1}{\kappa}\bigl[X_1 V^{-2} - X_2 V^{- (2
N_3 + 6)/3N_3}\bigr], \label{endenvi}
\end{equation}
where $$X_1 = \biggl[\frac{1}{3} - 3\frac{m^2 - mn + n^2}{N_3^2
(m+n)^2}\biggr]C_0,\quad X_2 = \frac{m^2 - mn + n^2}{N_0^2} -
\frac{X_1 A_1}{C_0}.$$ Further we obtain
\begin{equation}
\omega = \frac{p}{\varepsilon}  = -\frac{X_1  - X_4 V^{(4 N_3 -
6)/3N_3}}{X_1 - X_2 V^{ (4 N_3 - 6)/3N_3}}, \label{EoSvi}
\end{equation}
where $$ X_4 = \frac{2N_3 - 3}{3 N_3} A_0 + \frac{mn}{N_0^2} -
\frac{X_1 A_1}{C_0}.$$

\begin{figure}[ht]
\centering
\includegraphics[height=70mm]{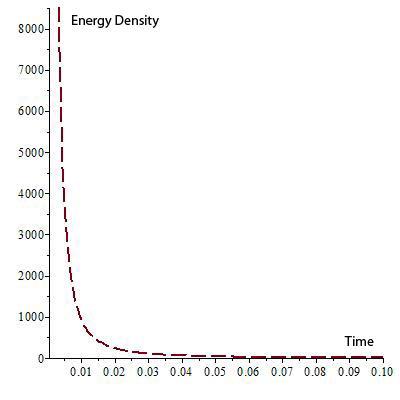} \\
\caption{Evolution of the energy density} \label{veE-6}.
\end{figure}

\begin{figure}[ht]
\centering
\includegraphics[height=70mm]{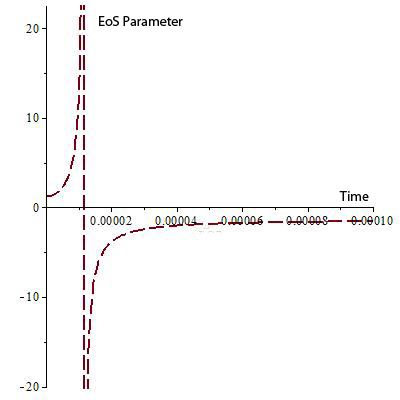} \\
\caption{Evolution of the EoS parameter. } \label{omE-6}.
\end{figure}

Figs. [\ref{veE-6}] and  [\ref{omE-6}] show the behavior of the
energy density and EoS parameter, respectively. As we see, energy
density is a decreasing function of time, while the EoS parameter
changes its sign.

So, if the present work is compared with experimental results
obtained in \cite{Knop,Tegmark1,Hinshaw,Komatsu}, then one can
conclude that the limit of $\omega$ provided by equation
\eqref{EoSvi} may accommodated with the acceptable range of EoS
parameter. Also it is observed that for $V = V_{c}$, $\omega$
vanishes, where $V_{c}$ is a critical Volume given by
\begin{equation}
 V_{c} = \biggl(\frac{X_1}{X_4}\biggr)^{3N_3/(4N_3 -6)}.\label{Vc}
\end{equation}
Thus, for this particular volume, our model represents a dusty
universe. We also note that the earlier real matter at $V \leq
V_{c}$, where $\omega \geq 0$ later on at $V > V_{c}$, where $\omega
< 0$ converted to the dark energy dominated phase of universe.

For the value of $\omega$ to be in consistent with observation
\cite{Knop}, we have the following general condition
\begin{equation}
 V_{1} < V < V_{2}, \label{V12vi}
\end{equation}
where
\begin{equation}
 V_{1} = \biggl(\frac{
X_1 + 1.67 X_1}{X_4 + 1.67 X_2}\biggr)^{3N_3/(4N_3 -6)},\label{V1vi}
\end{equation}
and
\begin{equation}
 V_{2} = \biggl(\frac{
X_1 + 0.62 X_1}{X_4 + 0.62 X_2}\biggr)^{3N_3/(4N_3 -6)}.\label{V2vi}
\end{equation}

For this constrain, we obtain  $-1.67 < \omega < -0.62$, which is in
good agreement with the limit obtained from observational results
coming from SNe Ia data \cite{Knop}.

For the value of $\omega$ to be in consistent with observation
\cite{Tegmark1}, we have the following general condition
\begin{equation}
V_{3} < V < V_{4}, \label{V34vi}
\end{equation}
where

\begin{equation}
 V_{3} = \biggl(\frac{
X_1 + 1.33 X_1}{X_4 + 1.33 X_2}\biggr)^{3N_3/(4N_3 -6)},\label{V3vi}
\end{equation}
and
\begin{equation}
 V_{4} = \biggl(\frac{
X_1 + 0.79 X_1}{X_4 + 0.79 X_2}\biggr)^{3N_3/(4N_3 -6)}.\label{V4vi}
\end{equation}

For this constrain, we obtain  $-1.33 < \omega < -0.79$, which is in
good agreement with the limit obtained from observational results
coming from SNe Ia data \cite{Tegmark1}.

For the value of $\omega$ to be in consistent with observation
\cite{Hinshaw,Komatsu}, we have the following general condition
\begin{equation}
V_{5} < V < V_{6},\label{V56vi}
\end{equation}
where
\begin{equation}
 V_{5} = \biggl(\frac{
X_1 + 1.44 X_1}{X_4 + 1.44 X_2}\biggr)^{3N_3/(4N_3 -6)},\label{V5vi}
\end{equation}
and
\begin{equation}
 V_{6} = \biggl(\frac{
X_3 + 0.92 X_1}{X_4 + 0.92 X_2}\biggr)^{3N_3/(4N_3 -6)}.\label{V6vi}
\end{equation}

For this constrain, we obtain  $-1.44 < \omega < -0.92$, which is in
good agreement with the limit obtained from observational results
coming from SNe Ia data \cite{Hinshaw,Komatsu}.

We also observed that if
\begin{equation}
 V_{0} = \biggl(\frac{
2X_1}{X_4 + X_2}\biggr)^{3N_3/(4N_3 -6)}.\label{V0vi}
\end{equation}
then for $V = V_0$ we have $\omega = -1$, i.e., we have universe
with cosmological constant. If $V < V_0$ the we have $\omega > -1$
that corresponds to quintessence, while for $V > V_0$ we have
$\omega
> -1$, i.e., Universe with phantom matter \cite{Caldwell1}.

From \eqref{endenvi} we found that the energy density is a
decreasing function of time and $\varepsilon \ge 0$ when

\begin{equation}
V \le \biggl(\frac{X_1}{X_2}\biggl)^{3N_3/(4N_3 - 6)}.
\label{deposvi}
\end{equation}

In absence of any curvature, matter energy density $\Omega_m$ and
dark energy density $\Omega_\Lambda$ are related by the equation

\begin{equation}
\Omega_m + \Omega_\Lambda = \frac{\varepsilon}{3 H^2} +
\frac{\Lambda}{3 H^2} = 1. \label{OmegamL}
\end{equation}

Inserting \eqref{Hubblevi1} and \eqref{endenvi} into \eqref{OmegamL}
we find the cosmological constant as
\begin{equation}
\Lambda = [3 C_0^2 - (X_1/\kappa)] V^{-2} + [3A_1 - X_2/\kappa]
V^{-2(N_3 + 3)/3N_3}, \label{Lambdavi}
\end{equation}
As we see, the cosmological function is a decreasing function of
time and it is always positive when

\begin{equation}
V \ge \biggl(\frac{X_1/\kappa - 3C_0}{3A_1 - X_2/\kappa
}\biggr)^{3N_3/(4N_3 - 6)}. \label{Lamposvi}
\end{equation}

\begin{figure}[ht]
\centering
\includegraphics[height=70mm]{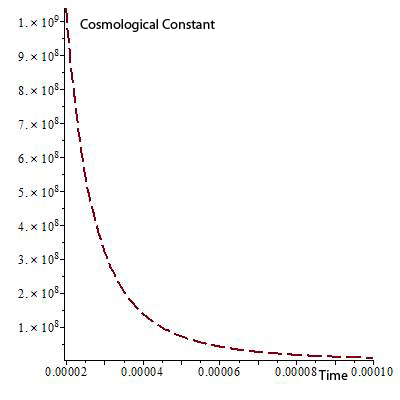} \\
\caption{Evolution of the cosmological constant} \label{LE-6}.
\end{figure}

Recent cosmological observations  suggest the existence of a
positive cosmological constant $\Lambda$ with the magnitude
$\Lambda(G\hbar/c^{3})\approx 10^{-123}$. These observations on
magnitude and red-shift of type Ia supernova suggest that our
universe may be an accelerating one with induced cosmological
density through the cosmological $\Lambda$-term. Thus, the nature of
$\Lambda$ in our derived DE model is supported by recent
observations. Fig. [\ref{LE-6}] shows the evolution of the
cosmological constant. As is seen, it is a decreasing function of
time.

For the stability of corresponding solutions, we should check that
our models are physically acceptable. For this, the velocity of
sound is less than that of light, i.e.,

\begin{equation}
0 \le v_s = \frac{dp}{d\varepsilon} < 1. \label{accon}
\end{equation}

\begin{figure}[ht]
\centering
\includegraphics[height=70mm]{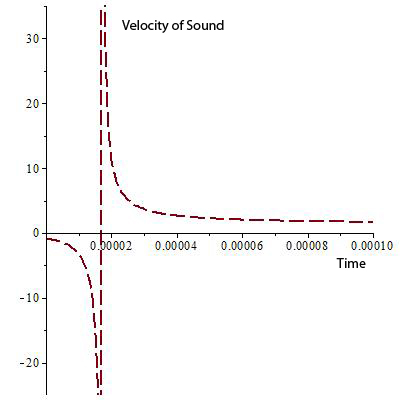} \\
\caption{Speed of sound with respect to cosmic time} \label{vsE-6}.
\end{figure}

In this case we find

\begin{equation}
v_s = \frac{dp}{d\varepsilon}  = -\frac{X_1  - [(N_3 + 3)X_4/3N_3]
V^{(4 N_3 - 6)/3N_3}}{X_1 - [(N_3 + 3)X_2/3N_3] V^{ (4 N_3 -
6)/3N_3}}. \label{vsvi}
\end{equation}
Fig. [\ref{vsE-6}] shows the behavior of $v_s$ in time.

As one sees, there are regions, where the solution is stable.
Choosing the problem parameters, such as $m,n,N_3$ we can obtain the
stable solutions.

\section{Conclusion}
In this report we have studied the evolution of the universe filled
with dark energy within the scope of a Bianchi type-VI model. In
case of a BVI model we found the exact solutions to the field
equations in quadrature. It was found that if the proportionality
condition is used, this together with the non-diagonal Einstein
equation leads to the isotropic distribution of energy momentum
tensor, i.e., $T_1^1 = T_2^2 = T_3^3$. This fact allows one to solve
the equation for volume scale $V$ exactly. The behavior of EoS
parameter $\omega$ is thoroughly studied. It is found that the
solution becomes stable as the Universe expands.


\begin{thebibliography}{999}

\bibitem{APSAPSSbi0}  Amirhashchi H., Pradhan A., and Saha B.,
Astrophys. Space Sci. {\bf 333} 295 (2011).

\bibitem{APSCPL}  Amirhashchi H., Pradhan A., and Saha B.,
Chinese Phys. Lett.  {\bf 3} 039801 (2011).

\bibitem {Hoogen}
Ib$\acute{a}\tilde{n}$ez J., van der Hoogen R.J., Coley A.A., Phys.
Rev. D {\bf 51} 928 (1995).

\bibitem{Caldwell1}
Caldwell R.R., Phys. Lett. B  {\bf 545}  23 (2002).

\bibitem{Hinshaw}
 Hinshaw, G., {\it et al.}, Astrophys. J. (Suppliment Series)
{\bf 180} 225 (2009).


\bibitem{kans66}  Kantowski R. and Sachs R.K., J. Math. Phys.
{\bf 7} 443 (1966).

\bibitem{ks66} Kristian J. and Sachs R.K.,  Apstrophys.
J. {\bf 143} 379 (1966).


\bibitem{Knop} Knop R.K.,
{\it et al.}, Astrophys. J. {\bf 598} 102 (2003).


\bibitem{Komatsu}
 Komatsu, E., {\it et al.},
Astrophys. J. (Suppliment Series) {\bf 180} 330 (2009).

\bibitem{PASIJTPbi}
Pradhan A., Amirhashchi H., and Saha B., Int. J. Theor. Phys. {\bf
50} 2923 (2011).


\bibitem{PASAPSS}
Pradhan A., Amirhashchi H., and Saha B., Astropys. Space Sci. {\bf
333} 343 (2011).


\bibitem{BVI} Saha B., Phys. Rev. D  {\bf 69} 124006 (2004).



\bibitem{GCBVI2010} Saha B.,
 Gravitation $\&$ Cosmology {\bf 16} 160 (2010).

\bibitem{SAPAPSS} Saha B., Amirhashchi H., and  Pradhan A.,
Astrophys. Space Sci. 2012 (online first)

\bibitem{svBVI}
Saha B. and Visinescu M., Romainan J. Phys. {\bf 55} 1064 (2010).

\bibitem {Socorro}
Socorro J., Medina E.R., Phys. Rev. D {\bf 61} 087702 (2000).

\bibitem{Tegmark1}
 Tegmark, M.,{\it et al.}, Phys. Rev. D {\bf 69} 103501 (2004).

\bibitem{thorne67} Thorne K.S., Astrophys. J. {\bf
148} 51 (1967).

\bibitem{Weaver1} Weaver M.,  Classical Quant. Grav. {\bf 17} 421 (2000).

\bibitem{YSAPSSbi}  Yadav A.K. and Saha B.,  Astrophys. Space Sci.
{\bf 337} 759 (2012).

\end{thebibliography}
\end{document}